\documentclass[journal]{IEEEtran}
\usepackage{flushend}
\usepackage{xcolor}
\usepackage{wrapfig}
\usepackage{cite}
\usepackage{graphicx}
\usepackage{epstopdf}
\ifCLASSINFOpdf
\else
\fi
\usepackage{amsmath}
\usepackage{textcomp}%for mu in title

\begin{document}
\title{4-dB Quadrature Squeezing with Fiber-coupled PPLN Ridge Waveguide Module}
\author{Naoto Takanashi, Takahiro Kashiwazaki, Takushi Kazama, Koji Enbutsu, Ryoichi Kasahara, Takeshi Umeki, and Akira Furusawa
\thanks{N. Takanashi and A. Furusawa are with Department of Applied Physics, School of Engineering, The University of Tokyo, 7-3-1 Hongo,Bunkyo-ku, Tokyo, 113-8656, Japan. A. Furusawa is the corresponding author to provide e-mail: akira@ap.t.u-tokyo.ac.jp.}
\thanks{T. Kashiwazaki, T. Kazama, K. Enbutsu, R. Kasahara, and T. Umeki are with NTT Device Technology Labs, NTT Corporation, 3-1, Morinosato Wakamiya, Atsugi, Kanagawa, 243-0198, Japan}
\thanks{This work is funded by Core Research for Evolutional Science and Technology (CREST) (JPMJCR15N5) of Japan Science and Technology Agency (JST), KAKENHI (18H05207) of Japan Society for the Promotion of Science (JSPS), APLS of Ministry of Education, Culture, Sports, Science and Technology (MEXT), and The University of Tokyo Foundation.}}

\markboth{Journal of Quantum Electronics, Letter}%~Vol.~XX, No.~X, January~2020}%
{Takanashi \MakeLowercase{\textit{et al.}}: Fiber-I/O Quadrature Squeezer showing 4-dB Noise Reduction based on a PPLN Ridge Waveguide}



\maketitle
%\IEEEpeerreviewmaketitle

\begin{abstract}
We have developed an optical parametric amplification module for quadrature squeezing with input and output ports coupled with optical fibers for both fundamental and second harmonic. The module consists of a periodically poled LiNbO${}_{3}$ ridge waveguide fabricated with dry etching, dichroic beamsplitters, lenses and four optical fiber pigtales. The high durability of the waveguide and the good separation of squeezed light from a pump beam by the dichroic beamsplitter enable us to inject intense continuous-wave pump light with the power of over 300 mW. We perform $-$4.0$\pm$0.1 dB of noise reduction for a vacuum state at 1553.3 nm  by using a fiber-optics-based measurement setup, which consists of a fiber-optic beamsplitter and a homemade fiber-receptacle balanced detector. The intrinsic loss of the squeezed vacuum in the module is estimated to be 25\%. Excluding the extrinsic loss of the measuremental system, the squeezing level in the output fiber of the module is estimated to be $-$5.7$\pm$0.1 dB. 
A modularized alignment-free fiber-coupled quadrature squeezer could help to realize quantum information processing with fiber optics.
\end{abstract}


% \fi
%\IEEEpeerreviewmaketitle

\section{Introduction}
\IEEEPARstart{Q}{uadrature} squeezed states are quantum resources in continuous-variable quantum information processing with quadrature amplitudes of light \cite{Squeeze:Walls,CVQI:PvL,EntanglementFromSq}. Especially, squeezed vacua are used as ancillary inputs for quantum operations such as quantum teleportation \cite{FurusawaTeleportation}, a quantum non-demolition gate \cite{QND:Shiozawa} and quantum key distribution \cite{KeydistributionSq}.

To realize a large-scale quantum circuit, it is important to utilize guided-wave optical components. Nowadays, with the development of telecommunication, highly reliable various fiber-coupled optical components such as lasers, modulators and optical beamsplitters have become commercially available. However, generation of high-level squeezed vacua still relies on optical parametric oscillators (OPOs) with bulk optics \cite{30yearsSq:Leuchs,7dBSq:Sasaki,15dBSq:Schnabel}. This is because guided-wave components has larger loss and lower durability for intense pump beams compared to free space optics. OPOs with capability of direct coupling with optical fibers \cite{Fiber:Fabre,takanashi} and a compact OPO on a breadboard \cite{CompactOPO} have been proposed but the need to control and adjust the cavity length could be an obstacle to scaling up of quantum circuits in the future.

By reducing propagation loss and improving the durability for intense pump beams, optical parametric amplification (OPA) on $\chi^{(2)}$ waveguides could take the place of OPOs. A practical advantage of OPAs is that, unlike OPOs, they do not require troublesome optical length control for the cavities. Additionally, OPAs can achieve THz-order operational bandwidth limited only by dispersion or phase matching conditions \cite{BroadbandOPA,FiberBroadbandOPA,WaveguideBroadbandOPA,OPAFurusawa}, while cavity structures of OPOs limit the bandwidth of the process. 
Although it is still below the level for quantum information processing, for example 3 dB, which is a condition of entanglement swapping \cite{3dBentanglementTheory}, the performance of fiber-coupled OPAs as sources of squeezed vacua have made remarkable progress in recent years. The level of squeezed vacua obtained from these components has reached 1.83 dB in 2016 \cite{fiberedOPA18dB} and 2.00 dB in 2019 \cite{fiberOPA20dB0}.

In this letter, we report generation and detection of a 4.0-dB squeezed vacuum from our newly developed fiber-coupled single-pass OPA module based on a dry etched periodically poled LiNbO${}_{3}$ (PPLN) waveguide. The module consists of the PPLN ridge waveguide, lenses for fiber coupling, four optical fiber pigtales, and dichroic beamsplitters. 
The high durability of the waveguide and the good separation of the dichroic beamsplitters allows to inject an intense pump beam, resulting high squeezing level. We detect $-$4.0$\pm$0.1 dB of squeezing and 14.1$\pm$0.1 dB of anti-squeezing at 10 MHz with pump power up to 330 mW. 
We use a fiber-optic beamsplitter for a homodyne detection, assuming applications of the module in fiber systems. Correcting for the deterioration caused by the measuremental system, the squeezing level and anti-squeezing level at the output port of the module are estimated to be $-$5.7$\pm$0.1 dB and 14.9$\pm$0.1 dB.
%%%%%%%%%%%%%%%%%%%%%%%%%%%%%%%%%%%%%%%%%%%%%%%%%%%%%%%%%%%%
\section{Device Design and Experimental Setup}
\begin{figure*}[!t]
\centering\includegraphics[width=13cm]{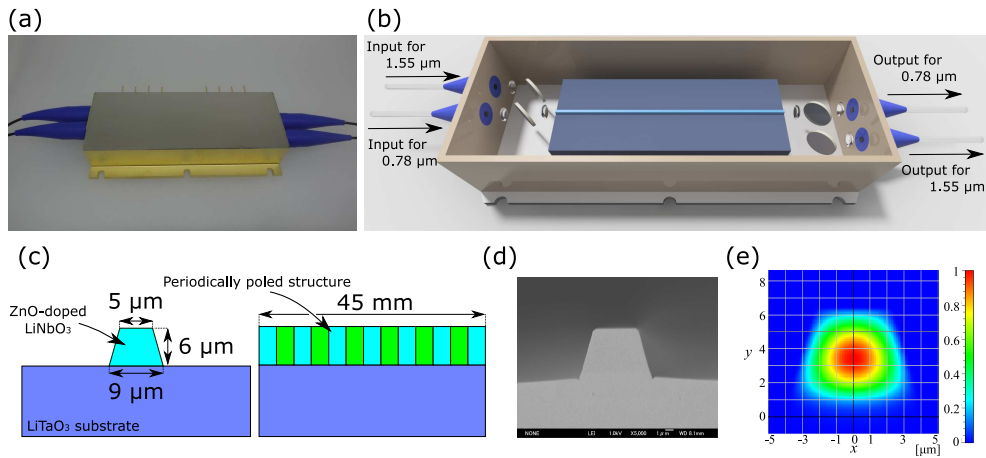}
\caption{Design of our OPA module. (a) Photograph of the module. (b) Schematic of the module. The module consists of a 45-mm long PPLN waveguide, four dichroic beamsplitters, six lenses and four pigtales. In this experiment, the input pigtale for a 1.55 $\mu$m beam is not used. (c) Schematic of the ridge waveguide. The substrate is LiTaO${}_{3}$, and the core is trapezoidal ZnO-doped LiNbO${}_3$. The waveguide has periodic poling for quasi-phase matching. (d) A picture of a waveguide end face taken with a scanning electron microscope. (e) A graph of the electric field amplitude distribution at the end face of the waveguide calculated by a computer simulation.}
\label{waveguide}
\end{figure*}
Figs. \ref{waveguide}(a) and (b) shows the external appearance and the internal schematic  of our OPA module. The module consists of a 45-mm long PPLN ridge waveguide, four dichroic beamsplitters, six collimation lenses and four pigtales. The temperature of the waveguide is controlled for quasi-phase matching. Dichroic beamsplitters (High transmission at 0.78 $\mu$m, high reflection at 1.5 $\mu$m) are used for separating a squeezed vacuum from a pump beam. For better separation, a squeezed vacuum is reflected twice on the beamsplitters. The good separation allows the intensity of a pump beam to be increased without any problems in the homodyne detection due to the transmission of an  intense pump beam. The transmittance of the module is 56\% at 1.55 $\mu$m and 60\% at 0.78 $\mu$m, which is mainly due to propagation loss in the waveguide and coupling loss on both ends. The propagation loss is considered to be mainly caused by surface roughness on the both sidewalls. The bandwidth of the parametric process in the PPLN waveguide is considered to be limited to THz order by its quasi-phase matching condition.

Fig. \ref{waveguide}(c) shows the schematic of our ridge waveguide. A core layer of PPLN is directly bonded to LiTaO${}_{3}$ substrate. The waveguide is fabricated by dry etching with argon gas using photoresist patterned by photolithography as an etching mask \cite{umeki-how2make}. Unlike diffusion methods, the direct bonding method does not cause defects in the crystal, which helps to increase the durability of the waveguide \cite{Kashiwazaki}. The dry etching method allows to make various shapes of waveguides. For instance, it enables to change the core size according to the position, while mechanical saws which are used to make diced waveguides moves only straightly. Taking this advantage, we create tapered structures at both ends of the waveguide to obtain better coupling with optical fibers. The tapered structures approximate the spot size of a propagating beam to a circle, which optimizes coupling efficiency with optical fibers under the constraint that the waveguide is trapezoidal.

Fig. \ref{waveguide}(d) is a picture of a waveguide end face taken with a scanning electron microscope and Fig. \ref{waveguide}(e) is a graph of the electric field amplitude distribution at the end face calculated by a simulation using a finite-difference method (Optiwave Systems Inc., OptiBPM 12). The mode-match between the calculated mode and a TEM${}_{00}$ mode is 98\%, although the actual coupling efficiency with the fiber is considered to be slightly lower due to manufacturing and assembling errors.
\begin{figure}[!t]
\centering\includegraphics[width=8.5cm]{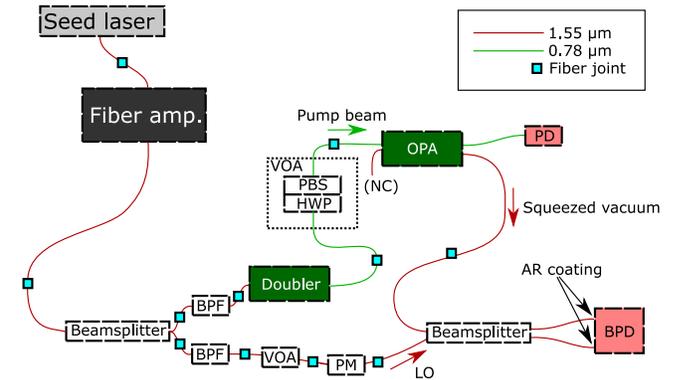}
\caption{Schematic of the experimental setup. A seed laser is a single frequency laser at 1553.3 nm. Output of the seed laser is amplified by a fiber amplifier. The amplified beam is split into two beams by a 10 dB (90:10) coupler, and the main output of the coupler pumps a frequency doubler after it passes through a bandpass filter. A frequency doubled beam pumps the OPA after it passes through a variable optical attenuator consisting of a half-wave plate and a polarizing beamsplitter. The intensity of the frequency doubled beam is monitored after transmission through the OPA. The tapped output of the 10 dB beamsplitter is used as a local oscillator of a homodyne measurement after passing a variable optical attenuator and a phase modulator. A squeezed vacuum from the OPA and the local oscillator are interfered in a 3 dB coupler, and the difference of the intensities of the output beams of the coupler is detected by a balanced photodetector. Note that only important elements are depicted. BPF, bandpass filter; VOA, variable optical attenuator; PBS, polarizing beamsplitter; HWP, half-wave plate; NC, not connected; PD, photodetector; PM, phase modulator; LO, local oscillator; AR, anti-reflection; BPD, balanced photodetector.}
\label{setup}
\end{figure}

Fig. \ref{setup} shows a schematic of the experiment. A source of continuous-wave laser light at 1553.3 nm is a narrow-linewidth and low-noise seed laser (RIO-OptaSense Inc., ORION module). The output of the seed laser is amplified by a Erbium-doped fiber amplifier (Keopsys, CEDA-C-PB-HP). The output of the fiber amplifier is splitted into two beams and passes through bandpass filters (Alnair Labs, TFF-15-1-PM-L-100-SS-SA) to reduce noise due to amplified spontaneous emission from the fiber amplifier.
The brighter output of the beamsplitter pumps a fiber-coupled frequency doubler (NTT Electronics, WH-0776-000-F-B-C). The frequency doubled beam passes through a custom-made fiber-pigtaled variable optical attenuator consisting of a half-wave plate (Casix, WPZ1210) and a polarizing beamsplitter (Sigma Koki, PBS-5-7800). This beam is used as the pump beam of the OPA module. The less intense output of the beamsplitter passes through a variable attenuator (Thorlabs, VOA50PM-FC) and used as a local oscillator (LO) for homodyne detection. The effect of phase noise of the fiber system is reduced by matching the optical length on the LO path with that on the path for generating the squeezed vacuum. As a result, the phase fluctuation is negligible during the measurement period.

The OPA module has two output ports. One is for 0.78 $\mu$m, and is used for monitoring the intensity of the pump beam. The intensity is measured by a Si photodetector (Newport 818-SL) and we estimate the incident pump power by dividing the monitored power by 0.6, which is the transmittance of the module at 0.78 $\mu$m. The other is for 1.5 $\mu$m, namely a port for the squeezed vacuum. The squeezed vacuum interferes with the LO in a 3 dB coupler (Thorlabs, PN1550R5F2). The phase of the LO is scanned by a phase modulator (Thorlabs, LN53-10-P-S-S-BNL). The output ports of the 3 dB coupler are spliced to anti-relection (AR) coated fibers (Thorlabs, P1-1550PMAR-2). The fibers are connected to a homemade fiber-receptacle InGaAs balanced photodetector consisting of lenses (Thorlabs, TC25FC-1550 and LA1134-C), mirrors (Sigma Koki, TFVM-25.4C05-1550), photodiodes (Laser Components, IGHQEX0100-1550-10-1.0-SPAR-TH-40) and an operational amplifier (Analog devices, AD829) with 18 k$\Omega$ of transimpedance. The signal from the detector is measured by a spectrum analyzer (Keysight, N9010A).

The spectrum analyzer is set to a zero-span mode at the measurement frequency of 10 MHz. The resolution and video bandwidths are set to 3 MHz and 510 Hz, respectively. The measurement frequency is the largest frequency without the deterioration of detector{\textquotesingle}s performance. The large resolution bandwidth and small video bandwidth help to obtain clear signal. Since the bandwidth of a single-pass OPA is on the order of terahertz, the frequency dependence of squeezing level is negligible at the order of megahertz.
%%%%%%%%%%%%%%%%%%%%%%%%%%%%%%%%%%%%%%%%%%%%%%%%%%%%%%%%%%%%%%%%%%%%%%%
\section{Result and discussion}
\begin{figure}[!t]
\centering\includegraphics[width=8cm]{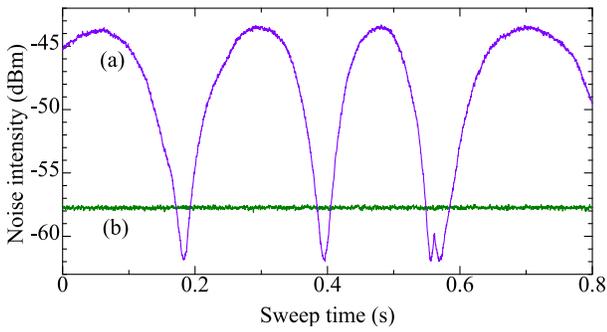}
\caption{Raw data of noise power as a function of the phase of the LO beam (scanned by a 1-Hz triangle wave). The symmetric structure around 0.56s is due to the reversal of the direction of scanning by the triangular wave. The intensity of a incident pump beam is %200 mW
330 mW%, measured at the output port of the module
. Center frequency is set to 10 MHz. Resolution bandwidth is set to 3 MHz and video bandwidth is set to 510 Hz. (a) Noise of a squeezed vacuum. (b) Shot noise.}
\label{200scan}
\end{figure}
\begin{figure}[!t]
\centering\includegraphics[width=8cm]{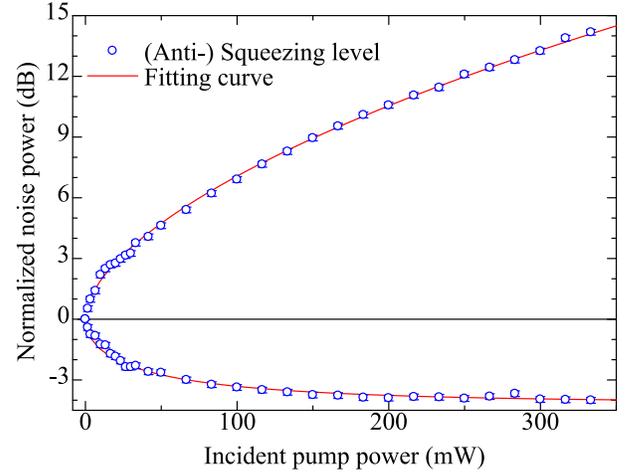}
\caption{Pump power dependency of the intensity of squeezed noise and anti-squeezed noise normalized by the intensity of shot noise. Note that the incident pump intensity is calculated by dividing the transmitted pump power measured at the output port of the module by 0.6, which is the transmittance of the module at 0.78 $\mu$m. Circles are measured values and curved line is a theoretical fitting.}
\label{pumpdep}
\end{figure}
Fig. \ref{200scan} shows the squeezed and anti-squeezed noise with the shot noise. The measurement frequency is set to be 10 MHz. The intensity of the pump beam is 330 mW. The transmittance of the module at 0.78 $\mu$m is 60\%. The intensity of the LO beam is 3.0 mW. 
The measured squeezing and the anti-squeezing level are $-$4.0$\pm$0.1 dB and 14.1$\pm$0.1 dB.

Fig. \ref{pumpdep} shows the pump power dependence of the squeezing and anti-squeezing level. The squeezing and anti-squeezing level $R_{\pm}$ with total detection loss $L$ are described as \cite{SqSHGPPLN}:
\begin{eqnarray}
R_{\pm} &=& L + (1-L) \exp(\pm 2\sqrt{ap}).
\end{eqnarray}
Here, $a$ is the efficiency of second harmonic generation. $L$, $a$ are fitted to be 38.6\%, 1034\% W${}^{-1}$, respectively.

To get the breakdown of the total detection loss $L$, transmittance of each element on the path of a squeezed vacuum is measured. The transmittance of the OPA module is measured to be 56\%, and that of the 3 dB coupler including a fiber joint loss is measured to be 45\%. Since the squeezed vacuum is generated inside the OPA module, assuming that a squeezed vacuum is generated in the middle of the waveguide, the effective loss can be considered to be $1-\sqrt{0.56}$, namely 25\%. For the 3 dB coupler, since the transmittance of a lossless coupler is 50\%, the effective loss is the excess loss of $1-0.45/0.50$, namely 10\%. The responsivity of the fiber-receptacle detector is measured to be 1.16 A/W, and it can be regarded as an effective loss of 7\%. The equivalent loss of the electric noise is 2 \%. Thus, total detection loss is calculated to be 38 \%, which is well-matched with the fitted value.

The coefficient $a$ represents the nonlinear efficiency of a waveguide. The fitted value is consistent with that of a similar waveguide, 1160 \% W${}^{-1}$ \cite{PPLNOPA6dB}.

Excluding the drop due to the detection efficiency, the original squeezing and anti-squeezing level at the output port of the module can be estimated to be $-$5.7$\pm$0.1 dB and 14.9$\pm$0.1 dB, respectively, which are consistent with those of a similar waveguide measured in a free-space setup \cite{PPLNOPA6dB}. The loss of a squeezed vacuum in the module is estimated as low as 25\%, which is considered to be mainly due to the propagation loss in the waveguide and the coupling mismatch with the output fiber. The propagation loss could be improved by improving dry etching method \cite{DryImprove} or performing wet etching after dry etching \cite{WetAfterDry}.
%%%%%%%%%%%%%%%%%%%%%%%%%%%%%%%%%%%%%%%%%%%%%%%%%%%%%%%%%%%%%%%%%%%%%%%
\section{Conclusion}
Measurement of a squeezed vacuum from a newly developed fiber-coupled single-pass OPA module was demonstrated in a fiber-optical setup. The PPLN ridge waveguide was fabricated with dry etching, which allows to fabricate a highly durable waveguide and to create a tapered structure at the ends of the waveguide to improve the coupling efficiency. The measured squeezing level is $-$4.0$\pm$0.1 dB, which is, to our knowledge, the best squeezing with fiber-coupled single-pass OPA to date. The module has input and output fibers for both fundamental and second harmonic. The good separation by dichroic beamsplitters and the high durability of the waveguide enable to inject an over-300-mW intense pump beam without any trouble in optical path for fundamental. We performed homodyne measurement with a fiber-optic beamsplitter and a fiber-receptacle balanced detector, looking toward fiber-optic applications. It is estimated that the original squeezing level at the output port of the module is $-$5.7$\pm$0.1 dB excluding the detection loss, which is consistent with that of a similar waveguide measured in a free-space setup \cite{PPLNOPA6dB}. A modularized alignment-free fiber-coupled squeezer with high-level noise reduction would play a important role in implementing quantum information processing with light in the near future.

\section*{Acknowledgments}
We thank Carlo Page, Taichi Yamashima and Asuka Inoue for feedback on the manuscript.

\bibliographystyle{IEEEtran}
\bibliography{example.bib}
\end{document}